\documentclass[twoside]{article}
\usepackage{epsfig,fleqn,espcrc2}

\usepackage{graphicx}
\usepackage[figuresright]{rotating}

\newcommand{\AmS}{{\protect\the\textfont2
  A\kern-.1667em\lower.5ex\hbox{M}\kern-.125emS}}

\def\refone{Ref.~\cite{CB-SChPT1}}

\def\chpt{\mbox{$\chi$PT}}
\def\figref#1{Fig.~\ref{fig:#1}}
\def\Figref#1{Figure~\ref{fig:#1}}
\def\half{{\scriptstyle \raise.2ex\hbox{${1\over2}$}}}
\def\fourth{{\scriptstyle \raise.2ex\hbox{${1\over4}$}}}
\def\eq#1{eq.~(\ref{eq:#1})}

\def\gtwid{\raise.3ex\hbox{$>$\kern-.75em\lower1ex\hbox{$\sim$}}}
\def\ltwid{\raise.3ex\hbox{$<$\kern-.75em\lower1ex\hbox{$\sim$}}}

\def\ie{{\it i.e.},\ }
\def\eg{{\it e.g.},\ }
\def\et{{\it et al.}}

\def\cL{{\cal L}}

\def\cO{{\cal O}}

\def\cV{{\cal V}}

\def\prd#1{Phys.\ Rev.\ {\bf D#1}}

\def\npb#1{Nucl.\ Phys.\ {\bf B#1}}

\def\boston{these proceedings}

\title{Chiral logs with staggered fermions\thanks{talk presented by C.\ Bernard at {\it Lattice 2002}; to be published in the proceedings}}

\author{C.~Aubin,\hskip-0.03in
\address{{\vskip-0.10in{\hskip 0.07in Department of Physics, Washington
University, St.~Louis, MO 63130, USA}}} 
C.~Bernard,$\null^{\rm a}$ 
\hskip-0.01in
C.~DeTar,\hskip-0.03in
\address{Physics Department, University of Utah, Salt Lake City, UT 84112, USA} 
Steven~Gottlieb,\hskip-0.03in
\address{Department of Physics, Indiana University, Bloomington, IN 47405, USA} 
Urs~M.~Heller,\hskip-0.03in
\address{CSIT, Florida State University, Tallahassee, FL 32306-4120, USA} 
K.~Orginos,\hskip-0.03in
\address{RIKEN BNL Research Center, Upton, New York 11973, USA} 
R.~Sugar\hskip0.005in
\address{Department of Physics, University of California, Santa Barbara, CA
93106, USA} 
and
D.~Toussaint\hskip0.005in
\address{Department of Physics, University of Arizona, Tucson, AZ 85721, USA} 
\advance\baselineskip -2pt
} 
       
\begin{document}

\begin{abstract}
We compute chiral logarithms in the presence
of ``taste'' symmetry breaking of staggered fermions.
The lagrangian 
of Lee and Sharpe is generalized and then used to calculate the
logs in $\pi$ and $K$ masses.
We correct an error in \refone;
the issue turns out to have implications for the
comparison with simulations, even at tree level.  
MILC data  with three light
dynamical flavors can be well fit by our formulas.  
However, 
two new chiral parameters, which describe $\cO(a^2)$ hairpin diagrams
for taste-nonsinglet mesons,
enter in the fits.  
To obtain precise
results for the physical $\cO(p^4)$ coefficients,
these new parameters 
will need to be bounded,
at least roughly. 
\vspace{-1pc}
\end{abstract}

\maketitle
It has become clear that simulation at rather light quark
mass is crucial for 
accurate determination of physical parameters, \eg
heavy-light decay constants \cite{KRONFELD}.  
Since staggered fermions provide the fastest known method of simulating
low-mass 
quarks, the systematic
effects associated with this fermion choice
should be studied.

We adopt the following nomenclature:
A single staggered field describes 4 equivalent
``tastes'' of quarks in the continuum limit;  taste symmetry is 
broken at $\cO(a^2)$.
We use the word ``flavor'' for
different staggered fields; lattice flavor symmetry is exact for equal masses.
MILC's improved staggered (``$a^2$-tad'') 
simulations \cite{MILC-RECENT} use three
fields (``$u,d,s$'') and reduce the tastes to one
per flavor by taking $\root 4 \of {\rm Det}$.
We call these ``$2+1$'' flavor simulations since we take
$m_u=m_d\equiv m_\ell$.

\begin{figure}[htb]
\null
\vspace{-.6truecm}
\includegraphics[bb = 0 0 4096 4096,
width=2.9truein]{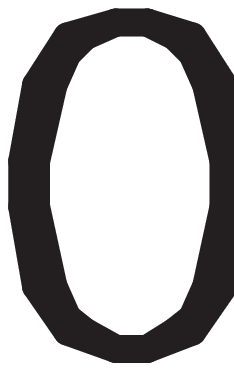}
\vspace{-1.25truecm}
\caption{\small
$m_{\pi,K}^2/(m_1+m_2)$ {\it vs.}
$m_1 + m_2$, in units of the potential scale $r_1$, for $2+1$ flavor lattices at $a\approx0.13$
fm.
The fit is to the continuum chiral log forms. The upper branch for
$(m_1+m_2)r_1\sim 0.2$ uses the kaon form; the rest
of the curve is the pion form.
}
\label{fig:nosplit}
\vspace{-0.28truein}
\end{figure}

MILC data for $m_\pi^2$ {\it vs.}\ quark mass
show clear deviations from linearity, as expected
from chiral logarithms.  
However, the detailed behavior at $a\approx0.13$ fm
 does not agree with continuum chiral perturbation 
theory (\chpt) \cite{GASSER-LEUTWYLER}.
\Figref{nosplit} shows a fit to the continuum forms
for $m_X^2/(m_1+m_2)$, where $X=\pi$ or $K$, and $m_1$, $m_2$
are quark masses.   
(We define ``pions'' to have $m_1=m_2$ and ``kaons'' 
to have $m_2\approx m_s^{\rm phys}$ and $m_1 \not= m_2$.)
The fit is very poor, with a confidence level of $5\times 10^{-5}$.

To get good fits, one needs to include the $\cO(a^2)$ taste-breaking effects
into the chiral calculations, \ie we need ``staggered chiral perturbation theory''
(S\chpt).
Such calculations start
with the chiral lagrangian of Lee and Sharpe \cite{LEE-SHARPE}
for 1 staggered flavor (4 tastes).  
One defines
\begin{equation}
 \Sigma(x) \equiv \exp(i\phi/f) \ ;\quad \phi = \phi_5 \xi_5 + \phi_\mu \xi_\mu + \cdots,
\end{equation}
where $\phi $ is a $4\times4$ matrix of pseudoscalar mesons of various tastes,
and $\xi_5,  \xi_\mu, \dots$ are taste matrices.
The Lee-Sharpe lagrangian is then
\begin{equation}
\cL= {f^2\over 8} {\rm tr}(\partial_\mu\Sigma
\partial_\mu \Sigma^\dagger)
-{\mu m f^2\over 4} {\rm tr}(\Sigma + \Sigma^\dagger) +  a^2\cV \  
\end{equation}
The taste-breaking potential $\cV$ has various operators with explicit taste matrices:
$ -\cV = \sum_{i=1}^6 C_i O_i$.
For example,
\begin{eqnarray}
O_2 & = & \half[{\rm tr}(\Sigma \Sigma) - {\rm tr} (\xi_5 \Sigma \xi_5 \Sigma) + {\rm h.c.}  ]\\
O_5 & = & \half[{\rm tr}(\xi_\nu\Sigma \xi_\nu\Sigma^\dagger) - {\rm tr} (\xi_\nu\xi_5 
\Sigma \xi_5 \xi_\nu \Sigma^\dagger)] \ .
\end{eqnarray}

To get NLO chiral results relevant to $2+1$ simulations, we follow a three step
procedure: 

{\small
1.\ Generalize the Lee-Sharpe lagrangian to
the 3-flavor case, \ie include 
three lattice staggered fields ($u, d, s$), each with
four tastes, and 
masses  $m_u=m_d\equiv m_\ell \not=m_s$.   
This  is an ``$8+4$'' theory.

2.\ Compute the desired quantities (here, $m_\pi^2$ and $m_K^2$) at one loop in the $8+4$ case.

3.\ Adjust the $8+4$ answer ``by hand'' to correspond to the $2+1$
case of interest by 
identifying the chiral contributions
that correspond to $n$ virtual quark loops
and dividing them by $4^n$.
}

Step 1 turns out to be non-trivial: \refone\ employed Fierz transformations
to simplify the flavor structure of the generalized potential $\cV$.
However, Ref.~\cite{LEE-SHARPE}
already used the Fierz freedom in deriving $\cV$.
The net result is that the generalized operators $O_2$ and $O_5$ are incorrect in
\refone.  

There are actually two operators that correspond
to each of $O_2$ and $O_5$:
\begin{eqnarray}
O_{21} & = & \fourth[{\rm Tr} (\xi_\mu \Sigma ){\rm Tr} (\xi_\mu \Sigma)  + {\rm h.c.}] 
\nonumber\\
O_{22} & = & \fourth[{\rm Tr} (\xi_\mu\xi_5 \Sigma ){\rm Tr} (\xi_5\xi_\mu \Sigma)  + 
   {\rm h.c.}] \nonumber \\
O_{51} & = & \half{\rm Tr} (\xi_\mu \Sigma ){\rm Tr} (\xi_\mu \Sigma^\dagger) \nonumber \\
O_{52} & = & \half{\rm Tr} (\xi_\mu\xi_5 \Sigma ){\rm Tr} (\xi_5\xi_\mu \Sigma^\dagger)   \ .
\label{eq:newops}
\end{eqnarray}
In the absence of flavor indices, the combinations
$O_{21}+O_{22}$ and $-O_{51}+O_{52}$ can be Fierzed into Lee-Sharpe $O_2$ and 
$O_5$, respectively; the other linear combinations turn into other $O_i$.  

The new operators have a surprising effect: they generate hairpin (disconnected)
diagrams for taste-nonsinglet (but flavor neutral) mesons.  
For taste vector and taste axial vector mesons,
this occurs at chiral tree level.
The magnitudes of the hairpins are 
\begin{equation}
\delta'_V \! \equiv\! {16a^2\over f^2}(C_{21}-C_{51});\;  \delta'_A \!  \equiv \!
\nonumber {16a^2\over f^2}(C_{22}-C_{52}) 
\end{equation}
for vector and axial taste, respectively.

The disconnected hairpin diagrams have not been included in any simulations to date
of taste-nonsinglet mesons.
Thus the  Lee-Sharpe lagrangian does not apply to such simulations:
The chiral lagrangian requires more than one flavor to describe particles,
\eg $\pi^+ = u\bar d$, that 
have no disconnected contributions.

At 1-loop, we need to iterate three kinds of hairpins on flavor-neutral
internal lines: the standard anomaly hairpin 
($m_0^2$) for taste singlets, and the
new hairpins for taste vector and axial vector.
Unlike  $m_0^2$, $\delta'_{V,A}$ 
cannot be taken to infinity.
We therefore have to rediagonalize the mass matrix in the taste $V, A$ channels.
We call, for example, the mass eigenstates in the flavor-neutral, 
taste vector channel $\pi^0_V$,
$\eta_V$ and $\eta'_V$.
Techniques in Ref.~\cite{SHARPE-SHORESH} are very useful
for reexpressing the iterated propagator as a sum of simple poles.

Adjustment to go from the $8+4$ to the $2+1$ theory is easy. Every
hairpin interaction on a meson line after the first introduces an additional
virtual quark loop.  Therefore, if $F(\Delta)$ is
the sum of hairpin diagrams, just take
$ F(\Delta) \to 4F({\Delta\over4})$ for $\Delta= \delta'_V$, $\delta'_A $ or $\delta$ 
(where $\delta\equiv m_0^2/(24\pi^2 f^2)$).

Additional diagrams that were 
identified as valence diagrams in \refone\ turn out to be
{\it disconnected} 
and  contribute only to flavor-neutral correlators.

We define   $\beta_{K^+_5}$ to be the chiral log term for the Goldstone $K^+$ mass
$(m^{\rm 1-loop}_{K^+_5})^2 /(\mu(m_\ell + m_s)) 
         =   1 + \beta_{K^+_5}/(16\pi^2f^2) + \cdots$,
and similarly for  $\beta_{\pi^+_5}$. 
%
Our results are then:
\begin{eqnarray}
& \beta_{\pi^+_5}  =   {2\delta'_V \over m^2_{\eta'_V}-m^2_{\eta_V}}
\Bigg[{m^2_{\eta'_V}-m^2_{S_V} \over m^2_{\eta'_V}-m^2_{\pi_V}}
m^2_{\eta'_V}\ln m^2_{\eta'_V} \nonumber \\
& \quad- {m^2_{\eta_V}-m^2_{S_V} \over m^2_{\eta_V}-m^2_{\pi_V}} 
m^2_{\eta_V}\ln m^2_{\eta_V} \Bigg] 
-4m_{\pi_V}^2 \ln m_{\pi_V}^2  \nonumber \\ 
& \quad+ ( V \to A)  
 + m_{\pi_I}^2 \ln m_{\pi_I}^2 - {1 \over 3} m_{\eta_I}^2\ln m_{\eta_I}^2 \nonumber \\
& \!\!\!\!  \beta_{K^+_5}  =  2\delta'_V
\Bigg[{m^2_{\eta'_V}\ln m^2_{\eta'_V}
-m^2_{\eta_V} \ln m^2_{\eta_V} \over m^2_{\eta'_V}-m^2_{\eta_V}}\Bigg] \nonumber \\
&\quad + ( V \to A)
 + {2 \over 3} m_{\eta_I}^2\ln m_{\eta_I}^2\ .
\label{eq:results}
\end{eqnarray}
Here $S$ is the $s\bar s$ meson.
The continuum result is in the taste-singlet channel ($I$); the remainder vanishes in
the limit $\delta'_{V,A}\to0$.

\begin{figure}[htb]
\null
\vspace{-.9truecm}
\includegraphics[bb = 0 0 4096 4096,
width=2.9truein]{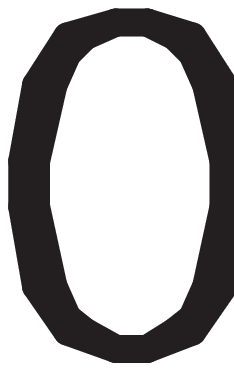}
\vspace{-1.2truecm}
\caption{\small Same as \protect{\figref{nosplit}}, but using the S\chpt\ forms,
\protect{\eq{results}}.  The parameter $\delta'_A$ is held fixed to a reasonably
small value in the fit.}
\label{fig:dmu5-.1}
\vspace{-0.30truein}
\end{figure}

With $\delta'_{V,A}$ as free parameters, these results give excellent fits to
the MILC data.
However, the parameters are
poorly constrained: $\delta'_V$ and $\delta'_A$ wander off in opposite
directions to large values (an order of magnitude larger than
known taste-violating terms in the $\pi^+$ sector).
On the other hand, if we fix $\delta'_V$ or $\delta'_A$ to be of reasonable
magnitude, the fits find reasonable values for the other parameter, too, and the confidence
levels are still excellent.
\Figref{dmu5-.1} shows the fit with $\delta'_A r_1^2$ fixed to $-0.1$;
$\delta'_V r_1^2$ is then found to be $0.25(11)$.
Based on the measured taste violations, the natural size for 
these parameters is $\delta' r_1^2\!\sim\!0.2$.

We conclude that
S\chpt\ is necessary to fit
current staggered lattice data.  The 
taste-violating hairpins introduce two new free parameters ($\delta'_{V,A}$)
into the chiral theory; other taste violations, which split
flavor-nonsinglet mesons, are not free parameters in the fits because
they are measured directly in simulations.  Unfortunately, 
the hairpin parameters are not well constrained, at least at present.
It does however appear to us that even a rough constraint on $\delta'_{V,A}$
(\eg demanding that they be no more than 3 times the known taste-violating terms)
will be sufficient to determine $p^4$ analytic coefficients (Gasser-Leutwyler ``$L_i$'')
with good precision.

We may be able to constrain the hairpin parameters in a variety of ways.
Additional S\chpt\ calculations are underway \cite{AUBIN-CB} that generalize these
results to the quenched and partially quenched cases, and extend them to pseudoscalar decay
constants (of both light-light and heavy-light mesons).
With these calculations  in hand,
the direct constraints from all the fits may prove sufficient.	
Another approach is to compute the four-quark, taste-violating operators 
perturbatively \cite{LEPAGE};
the  hairpin parameters may then be estimated by vacuum saturation or lattice calculation
of the matrix elements.
Finally, a direct lattice evaluation of the disconnected hairpin graphs may be possible.

We are grateful to M.\ Golterman, G.\ P.\ Lepage, and S.\ Sharpe for very useful discussions.
Computations were performed at LANL, NERSC, NCSA, ORNL, PSC and SDSC. This work
was supported by the U.S.\ DOE and NSF.

\def\refone{\cite{CB-SChPT1}}

\end{document}